\documentclass[manuscript]{aastex}
\begin{document}
\title{Three Fundamental Periods in a 87 Years Light Curve of the Symbiotic Star MWC 560.}
\author{Elia M. Leibowitz\altaffilmark{1} and Liliana Formiggini\altaffilmark{1}}
\affil{The Wise Observatory and the School of Physics and Astronomy, Raymond
and Beverly Sackler Faculty of Exact Sciences \\ Tel Aviv University, Tel
Aviv 69978, Israel\\}

\email{elia@astro.tau.ac.il}
\begin{abstract}

We have constructed a visual light curve of the symbiotic star MWC covering the last 87 years of its history. The data were assembled from the literature and from the AAVSO data bank. Most of the periodic components of the system brightness variation can be accounted for by the operation of 3 basic clocks of the periods P1=19000 d, P2=1943 d and P3=722 d. These periods can plausibly, and consistently with the observations, be attributed to 3 physical mechanisms in the system. They are, respectively, the working of a solar-like magnetic dynamo cycle in the outer layers of the giant star of the system, the binary orbit cycle and the sidereal rotation cycle of the giant star. MWC 560 is the $7^{th}$ symbiotic star with historical light curves that reveal similar basic characteristics of the systems. The light curves of all these stars are well interpreted on the basis of current understanding of the physical processes that are the major sources of the optical luminosity of these symbiotic systems.

\end{abstract}


\section{Introduction}

The remarkable symbiotic system MWC 560 (V 694 Mon) was discovered by Merrill and Burwell (1943) as an object with bright hydrogen lines.
The blue continuum and the observed TiO bands led to its classification as a symbiotic system (Sanduleak \& Stephenson 1973). It was quite neglected by observers until early 1990 when a 2 magnitude  rising in brightness brought it back into attention (Tomov 1990). The long-term light variations show a period of about 1930 days (Doroshenko et al. 1993) assumed to be the binary orbital period of the system. It is commonly believed that the system is a binary one consisting of an M type giant and a white dwarf with an accretion disk around it. It has been also suggested that the ~1930 d period is that of a precession of an accretion disk around the hot component (Doroshenko et al. 1993, Iijima 2002), or that it is the period of pulsation of the M giant of the system (Doroshenko et al. 1993).

Its optical spectrum shows highly variable absorption features. The blue shifted ones are interpreted as due to a jet outflow along the line of sight (Tomov et al. 1990). Jets and jet outflows are known to exist in several symbiotic system (Leedjarv 2003), but only for MWC 560 the jet axis is nearly parallel to the line of sight to the star (Schmid et al. 2001).
Short term flickering activity  with quasi periodic modulation from
minutes to hours was detected during and after the outburst of 1990 (Tomov et al. 1990, 1996; Michalitsianos et al. 1991,1993, Dobrzycka et al. 1996, Zamanov et al. 2011). 

More recently Gromadzki et al. (2007) analyzed the data from 1990 to 2007.
Beside the main period of 1931 d, they found a period of 340 days in
J,H,K,L photometry. Frackowiak et al. (2003) announced a discovery of 
a 161.3 d periodicity in the LC of the system which they have attributed to 
pulsations of the giant star.

In this paper we collected all the available photometric data of MWC 560 from 1928 to 2015 with the aim of analyzing the long-term photometric behaviour of this star.

\section{The data}

Luthardt (1991) published a light curve (LC) of MWC 560 obtained from 750 plates of the
Sonnenberg sky patrol, covering the period from 1928 to the beginning of 1990. Tomov et al. (1990, 1996) obtained photoelectric data during and after the outburst of 1990. According to Doroshenko et al. (1993)  a shift of 9.44 m is needed to transform plate magnitudes of Luthardt data to the photoelectric system of Tomov and of his own observations. Visual observations by the American Association of Variable Stars Observers are available in the archive of the organization from April 1990 up to the present day. Most of the photoelectric measurements of Tomov have been performed in a time interval JD 2447985-2448385,  already covered by the AAVSO data. We found that  the two data sets along their common time interval, at the level of accuracy of 0.05-0.1 magnitude, which is the level of our interest in this work,  are virtually the same.   

With the correction factor suggested by Doroshenko et al. 1993, we were thus able to connect all the data at hand to magnitudes on a single common scale, the visual magnitude scale of the published AAVSO data.

\section{Analysis}

\subsection{Full Light Curve}

Figure 1 is the visual light curve of MWC 560, constructed as described in the previous section. It covers a time interval of 31561 days, from 1928 September 27 until 2015 February 23, JD 2425516-2457077.  There is a distinct division between 2 different sections in the LC, 
to which we refer as sections A and B.  In April/May 1989, around JD 2447700, 
the mean magnitude of the star, over a few hundred days, has brightened by about  1.2 
magnitude. By coincidence, the two sections are also clearly different in the data 
sources that established them. Section A consists mainly on magnitudes measured on historical photographic plates while Section B is mainly AAVSO visual measurements by members of this organization  The AAVSO data have been binned into 10 days wide bins.

The upward step in a presumably DC component of the LC of the star introduces spurious peaks in the presentation of the LC in the Fourier space. Taking the power spectrum (PS) of the entire LC would therefore be a poor technique for searching for genuine periodic components in the LC. In order to inquire whether or not the LC includes periodic oscillations that are coherent throughout the whole duration of the measurements we created an overall pseudo-observed LC by subtracting from each of the two sections the mean magnitude of the star over the time duration of the respective section. This procedure amounts to an assumption that the apparent oscillations of the star are superposed on a DC component of nearly steady brightness that underwent a discreet event, of a time scale of a few tens of days, of a stepwise increase in its brightness. The LC with the 0 mean magnitude of its two sections was then detrended by subtracting from it a $2^{nd}$ degree polynomial fitted to the data points by a least squares procedure. We refer to the resulting time series as the Z' LC. 

As seen in Figure 1, the density of the data points on the time axis in section B is much larger than in section A. In order not to give section B an overdue statistical weight in the analysis of the Z'  LC we dilute the number of points in it so as to have a uniform density of points over the entire time axis. This was done by randomly selecting from the B section a number of points, eliminating all the others, such that the mean density of the remaining points over the B time axis will be equal to the density of points in the A section. The resulting series is a Z LC and one such LC is presented as dots in Figure 2(a).  

Each process of the random selection of points from the B section naturally creates a slightly different realization of the Z LC. We find that results of our interest from the analysis of the Z LC are insensitive to the particular realization of the Z series. The dots in Figure 2(a) display one realization of the Z LC. The solid line in this figure will be explained in Section 4.1. We refer to magnitudes of the Z LC as Yz points.

The solid line in Figure 2(b) presents the PS of  the Z LC shown in frame a, in the frequency range corresponding to the periods interval 25000 to 250 d. In the higher frequency range corresponding to periods down to 100 d, not displayed in the figure, there are no peaks with maximum power greater than 6,  considerably lower than the horizontal line in the figure.  This line represents the limit of a false alarm probability  smaller than $e^{-3}$. This means that only peaks in the PS that are higher than this line have a probability $e^{-3}$ or smaller that their appearance in the PS is due to random noise in the data. In drawing this line we used expression (18)  in Scargle (1982). The lower curve will be discussed in Section 4.1.
 
The two highly significant peaks in the upper curve in Figure 2(b) are at the frequencies corresponding to the periods  9651 d and 1947 d.  By least squares search of best fitted pair of periods to many different Z LCs we find for the mean best fit periods the values P = 9500 $\pm$ 700 d and P2 = 1943 $\pm$ 12 d. The error estimates are based on the dispersion in the resulting pair values in the ensemble of the Z LCs that we have tested.  Figures 2(c) and 2(d) display the folding of the Z Lc presented in frame a onto the two periods P and P2. The cycle is displayed twice. For reasons specified below, we consider the period P as one half of the period P1=2P=19000 d.
 
The period $\sim$ 1943 d has been recognized in the LC of this star already in the past by a number of investigators (Doroshenko et al. 1993, Gromadzki et al. 2007). Our present analysis demonstrates that it is a coherent oscillation of the star throughout the 31561 days covered by the observations. It preserves the same phasing in the two sections of the LC in spite of the large difference of $\sim$1.2  in their mean magnitude. This conclusion is also demonstrated visibly in Figure 2(c).  Figure 2(d)  shows in the same way that t...he 9500 periodicity is also preserving its phase throughout the entire LC duration. We note however that while the LC covers more than 16 cycles of P2, it covers only 3.3 cycles of P. Therefore the question whether or not P is a persistent periodicity of the system must remain open. We shall return to this question in  section 5.2 of this paper.

\subsection{Section A}

Figure 3(a) presents the LC of Section A.  The dots display the detrended series by subtracting from the LC a least squares best fitted polynomial of $2^{nd}$ degree. It covers the time interval of 22194 days from September 1928 to May 1989, JD 2425516-2447710. We denote this series as the Ya function. The solid line will be discussed in Section 4.1.

The upper solid curve in Figure 3(b) displays the PS of Ya, in the frequency interval corresponding to the period interval 22194 - 250 d. The horizontal line is as in Figure 2(b). All peaks in the PS of Section A at higher frequencies up to a the frequency  1/100 $d^{-1}$, which is a mean Nyquist frequency of this time series, are well below the horizontal line in the figure.

The lower, red line in the figure will be explained in Section  4.1.

\subsection{Section B}

Figure 4(a) presents the LC of Section B. It covers  9164 days from May 1989 up to February 2015, JD= 2447908-2457077. The dots are the detrended series by subtracting the linear trend from the data.  We denote this LC as the Yb function. The solid line will be discussed in Section 4.1. 

The upper solid curve in frame b is the PS of the Yb LC. The period range covered in the figure is 9164-250 d. The horizontal line is as in Figure 3(b). Here again, at all higher frequencies not shown in the figure up to some mean Nyquist frequency of the series, there are no peaks higher than the horizontal line.

The lower red line will be explained in Section 4.1.

\section{The Temporal Content in the LC}
\subsection{The 8 basic periods}

Examination of periods corresponding to the highest peaks in the 3 PSa presented in Figures 2(b), 3(b) and 4(b) led us to suggest that  most of the periodic content of the 87 years LC of MWC 560 can be accounted for by an interplay among 3 basic independent periodicities. They are P1=19000,  P2=1943 and P4=1150 d, with P1 that is present in the LC by its first 6 harmonics. We preserve the notation P3 for another periodicity to be introduced in section 5.4. We therefore consider the following vector of periods

V=[P1, P2, P4, P1/2,P1/3,P1/4,P1/5,P1/6]=[19000,1943,1150,9500,6333,4750,3800,3167] days.

We construct a synthetic light curve Ysa by computing the value of a series of 8 harmonic waves of the periods of vector V, fitting it by least squares to the observed LC Ya. Because of the much shorter duration of section B, fitting this series of 8 different periods to it produces unrealistic large amplitudes of some of the series components. Therefore, in constructing the model Ysb for Yb we consider a 6 term series of the periods vector $V^{(-)}$ which is  the V vector, from which the component P1 and P1/3 have been omitted.  Note that in the analysis of the LC of section B alone, the full Yb data, shown on the right hand side of Figure 1, were used, without applying the dilution operation.

The solid curves in Figures 3(a) and 4(a)  present the theoretical LCs so constructed as models for the observed ones Ya and Yb. The lower red curves in Figures 3(b) and 4(b) are the PSa of these theoretical LCs.

As mentioned above and seen in Figure 1,  in April/May 1989 the stellar system MWC 560 has undergone a dramatic event of a sudden tripling  its optical luminosity. In analyzing the full LC of the system we consider 2 possibilities.  (I)  The 1989 event did not affect the coherence of the periodicities that underline the star periodic behavior. (II)  The 1989 event has not changed the periodicities but for some of them the event was associated with a measurable phase shift. 

For modeling Yz in case I we fitted the series of the harmonic waves of the V vector denoted YszI to the Yz LC shown as dots in Figure 2(a).  To take into account the case II possibility, we created a model for Yz, denoted YszII by combining Ysa, the synthetic LC for Ya, with Ysb, the synthetic LC for Yb. The combination was conducted in the same procedure which we applied in the creation the Yz LC from the observed Ya and Yb LCs,  including the dilution of the function Ysb, as explained in section 3.1.

The solid curve in Figure 2(a) is the synthetic LC constructed for case II and the lower red  curve in Figure 2(b) is its PS. The corresponding curves for case I are nearly indistinguishable by eye from those of case II shown in the figure.

\subsection{Reality of the model}

The apparent good fit of the synthetic LCs and their PSa to the observed LCs  and the corresponding PSa as presented in Figures 2, 3 and 4 is not too surprising. The fitted synthetic functions have quite a large  number of adjustable parameters, namely, the amplitude and phase of each of the 8 periods considered, or the 6 in the case of Yb. In this section we argue that notwithstanding this limitation, our suggestion that the 3 periods P1, P2 and P4 are truly dominating the long term variability of the star is well founded.

First we note that 5 out of the 8 periods considered are not independent of the other 3 but are harmonics of P1. Secondly, the very same periods are fitted both to the Yz and to the Ya functions, albeit that their PSa are not identical.  The 6 periods of the series fitted to the Yb LC, the PS of which is not identical to either one of these two, are also a subgroup of the same 8 periods fitted to the other LCs.  

Our main argument for the reality of our suggestion is based on the fact that the LCs constructed out of the 3 basic periods as described above fit the observed LCs better than harmonic series of similar or greater number of independent periods, the values of which are taken directly from the PSa of the observed LCs themselves. 

Our second major evidence lies in the fact that the periods of a number of high peaks that stand out in the observed PSa are beats of pairs of periods from our suggested triple basic ones.

\subsection{Fit Quality}

We measure the quality of a fit of a model to an observed LC by the value of the parameter S$_{lc}$, the standard deviation (StD) of the difference between observed and theoretical values on all times of the observations. Similarly we use the parameter S$_{ps}$, the StD of the differences between the PSa of the corresponding LCs, as a measure of the fit of the PS of the theoretical LC to that of the observed LC. We note already at this stage, that  S$_{lc}$ measures the fitting of details of the LC, in particular it takes into account differences between the theoretical and the observed LCs on all time scales of the light variations, including those that are due to all sources of noise. In contrast, S$_{ps}$ is a measure of the fit of the major underlying harmonic content of the model to that of the observed LC.

As reference for the quality of our suggested synthetic LC Ysa we have considered another series of 8 harmonic waves with the periods corresponding to the 8 highest peaks in the 'clean' PS of the observed LC Ya, which we denote as Yta. 

In 'clean' PS we mean a PS from which peaks associated with periods that are merely aliases of periods corresponding to higher peaks in the PS, due to the unequal spacing between the observations on the time axis, are eliminated. This is performed by a computer program that we have developed in the following way. For the period of each high peak in the PS of the observed LC, beginning with the highest one, the routine fits an harmonic wave of that period to the observed LC by the least squares procedure. High peaks in the PS of this time series associated with periods others than the period of the generated wave itself, are considered aliases and their counterparts in the PS of the observed LC are eliminated. Our program is essentially a variation on the theme of the 'CLEAN' type procedures that are much in use the astronomical literature (Roberts et al. 1987)

As reference for the fit quality of our synthetic LC Ysb we constructed in a similar way a series of 6 harmonic waves of the 6 highest 'clean' peaks in the PS of Yb. Another synthetic LC, YtzI, is an 8 term series of the 8 highest peaks in the 'clean' PS of the observed LC Yz. Note that the 8 periods taken from the PS of Ya and used in the construction of Yta are not identical to the 8 periods that are fitted to Yz. Similarly, the 6 periods of which Ysb is created are not subgroup of the 8 periods of Ya, neither are they subgroup of the 8 periods of Ysz. This is in contrast to the relation among the periods at the basis of our suggested models for Ya, Yb and Yz, that are all the same, or among the same 8 periodicities. Thus, in addition to the fact that the periods used in the reference LCs are drawn directly from the observations, the number of adjustable parameters in the reference LCs is much higher. 

Table 1 presents the values of the 2 fitting parameters S$_{lc}$ and S$_{ps}$ for these 6 synthetic LCs that we have constructed. In the two left columns one sees that the PS of our suggested model for the Ya LC fits the PS of the real data better than a series with the same number of terms, built upon the peak periods of the PS of the real data. The S$_{lc}$ number shows that the fit of our model LC  to the observed one, in the time domain is slightly worse than the fit of the reference artificial LC Yta which uses peak periods derived from the real data. 

The middle 2 columns of the table show that our suggested model for Yb fits the observed data less well than the artificial LC built upon observed periods. The rightmost two columns show that under the case I assumption,  in its real time presentation as well as  in its approximate presentation in the Fourier space as manifested by the PS function, our suggested model fits the real data better than the artificial LC built upon periods derived directly from the data.

For the analysis of case II, for comparison with our model YszII we have created two additional artificial LCs that are combination of LCs that are fitted to the observed Ya and Yb LCs separately. The first one,  YtzII, is the combination of artificial LCs Yta and Ytb performed in the process described above. The second one YfzII is  an 8 terms series of the periods corresponding to the 8 highest peaks of the 'clean' PS of Yz fitted to Ya, combined with a 6 term series of the same 8 periods, from which the largest and the third largest periods were omitted, fitted to the Yb function.

Table 2 presents the Std values for these 3 artificial LCs that we have so constructed for case II. 

Here one should compare the numbers in the $3^{d}$  column to those in the first and the $2^{nd}$ columns. One sees the our model fits the observed LC and its PS  better than a combination of the fit of the observed peak periods, fitted independently to Ya and Yb. One sees further, that even when we combine the fit of the observed period of Ya, with the observed periods of Yb which are different from the Ya periods, our model still produces a better fit to the observed PS of Yz, and only very slightly higher value of the S$_{lc}$ parameter that measures the fit to the LC itself.

Thus, the PSa of all our 8 periods models, except for the short LC of Section B, fit the PSa of the corresponding observed LCs better than the PSa of LCs that are built upon similar or larger number of periods derived directly from the 'clean' PSa of the observed data.

\subsection{Beats}

The  $1^{st}$  column of Table 3 lists all 6 possible beats among the three periods P1, P2 and P4 that we are suggesting to be underlying the variability of the MWC 560 stellar system in the last 87 years. The $2^{nd}$ column lists the corresponding period values. In the last 3 columns we present period values of clearly distinguishable peaks in the PSa of the corresponding LCs listed at the head of each of these columns. These peaks are marked in figures 2(b), 3(b) and 4(b) by arrows and the corresponding peak period values, presented without brackets. The similarity between any number presented in either one of the last 3 columns of the table,  and the number in column 2 on the same line,  can hardly be ignored.

The 2 marked peaks in Figure 4(b) rise above the horizontal line, indicating that formally they are statistically significant at the 95 percent confidence level. The marked peaks in Figure 2(b) and 3(b) are all beneath the horizontal line, meaning that each one of them by itself cannot be regarded a statistically significant period of the system. However, in their entirety, all the marked peaks are among the most pronounced ones in the PSa in which they appear, some of them are among the periods considered in constructing the various Yt reference LCs. 

In the lower curves in Figures 2(b), 3(b) and 4(b) we see that in the PS of our model LCs, there are no peaks corresponding to the ones in the upper curve marked by the numbers without brackets.  The small feature flanking on both sides of the 1150 d peak in the lower curve of Figure 3(b) are of much lower height than the marked peaks above them and their peak periods are different, albeit slightly, from the peak periods in the observed PS. Similarly the 2 apparent weak features seen in the lower curve of Figure 4(b) underneath the marked peaks  1244 and 1074 in the upper curve are much lower than the observed peaks and their peak period do not coincide with the indicated numbers in the figure.

The absence of the marked peaks in the upper curves from the corresponding lower curves is of course not surprising since we have not introduced these periodicities into our Ys LCs. The marked peaks in the upper curves must therefore be genuine features of the observed LC and are not artifact of an interference among the  periods that we are considering. 

Each one of the above mentioned agreements between numbers in the 3 last columns of Table 3 and the numbers in column 2 on the corresponding same line may be a random coincidence. However the overall number of all these agreements, always with beats of two periods out of the same triple P1, P2 and P4 periods, is unlikely to be a random coincidence. We believe that it gives credence to our claim that the triple periods are indeed real characteristics of the MWC 560 stellar system.

Note also that the 8 periods that we are considering in our suggested models do not include any of the marked beat periods which are of noticeable signal in the observed LCs as evident by the distinguishable features in the corresponding PSa. Nevertheless the general fit of the PSa of our models, excluding that for Yb, are still better than the fit of the Yt functions.

\subsection{Coherence}

As described in section 4, we constructed the synthetic LCs Ysa and Ysz by least squares fitting of  series of 8 waves of the 8 periods of the period vector V, to the observed  Ya and Yz LCs. We eliminated the periods P1 and P1/3 from the vector V and fitted a series of 6 waves of the $V^{(-)}$ vector of the remaining 6 periods to the observed Yb LC. In the resulting Ysb LC, the amplitude of the P4 wave is negligibly small. Indeed, a glance in Figure 4(b) reveals that there is no discernable peak at this periodicity in the PS of Yb.

Column 1 of Table 4 lists the values of the 5 input periods that are common in the establishment of the Ysa and Ysb LCs, for which the best fit procedure yields appreciate amplitudes in both synthetic LCs. For these periods the resulting phases from the least squares procedure is therefore meaningful. 

Column 2 of Table 4 presents for each of these 5 periods the difference in phase between the 8 terms fit to Ya and the 6 terms fit to Yb. As explained in section 4.1, these fittings are performed independently of each other.

Column 3 of Table 4 lists the 5 high peak period values, as determined by the 'clean' PS of the observed Yz LC, that correspond to our 5 input periods listed in column 1. Column 4 presents the difference in phase of the corresponding best fitted waves, between the fit of the 8 'clean' periods of Yz to the LC  Ya, and the 6 'clean' periods of Yz fitted to Yb.  

Comparing column 2 and column 4 of Table 4 we observe that there is no shift in the phase of P1/2 and P1/4 between section A and B, while with the corresponding 'clean' periods 9507 and 4664 the phases between the two sections of the LC do not match. 

The higher harmonics of P1 are more contaminated by higher frequency oscillations of the star and/or by high frequency noise fluctuations in this observed LC which is of limited time duration.   

\subsection{Additional periods}

\subsubsection{The $\sim$ 1940 periodicity}

Gromadzki et al. (2007) have conducted a period analysis of the visual LC of MWC 560, as well as of  LCs of a few IR bands in the radiation of the system. They proposed a period Pd=1931 d in the visual LC, a value that is similar to the one suggested also by Doroshenko et al. (1993). In Section 3 we have derived from a much longer LC of the star the value P2=1943 d. We believe that our value for this periodicity is slightly nearer the true period in the LC of the star.

In Section 3 we showed that a few notable features in the PSa of the observed LCs, marked in Figures 2(b), 3(b) and 4(b), appear at periods that are beats of the period P2 with P1 or P4. If Pd is taken as the period beating with P1 and P4 instead of P2, the periods of the marked features in the PSa do not fit the resulting beat periodicities as well as when P2 is taken as the beating period. 

The last line of Table 4 shows that when fitting the P2 period to section A and section B of the LC, along with the other 7 periods of our suggested model, the resulting P2 wave is found to be coherent throughout the entire 87 years of observations in this star. In contrast, when the period Pd is considered, along with the other 7 periods corresponding to the highest peaks in the PS of the Yz LC, there is a difference of 0.06 in the phase of the Pd wave between the two sections of the Yz LC. 

We have also fitted a single sine wave of period Pd to section A of our LC and independently also to section B of the LC. There is a shift of 0.1 in the phase between these 2 waves. When performing the same calculations with the P2 periodicity, the phase shift found between the two waves is only 0.06. 

We consider the better coherence of the P2 wave throughout the whole time interval covered by the observations as a further evidence in favor of the value P2 over Pd as the periodicity of this order in the light of the star. We suggest the following ephemeris for maximum light of this cycle: 

\begin{equation}     
 max light= JD 2455799+1943*E.
\end{equation}
The ephemeris for maximum light in the P1/2 cycle is:
\begin{equation}
 max light= JD  2456536+9500*E .                                                       
\end{equation}

\subsection{Claimed Periodicities}
\subsubsection{The 339 and 308 periodicities}

Gromadzki et al. (2007) suggested also the period Pp=339 d as another characteristic of the system, possibly the period of pulsations of the cool component. They also raised the possibility that Pq=308 d is another characteristic of the system. We also find traces of features at these periods in PSa of some of the LCs that we have considered but we regard it rather doubtful that these two periods are real characteristics of the MWC 560 stellar system.

As regards to the 308 period, Gromadzki et al. (2007) themselves drew attention to the possibility that it is an alias of the P2 period due to the 1 year cycle in the distribution of the observational points on the time axis.
 Indeed we have: $1/(1/365+1/1943) = 307.28$.

In Figures 2(b), 3(b) and 4(b), a peak corresponding to the 308 d period in the PS of the observed LCs in marked with the number in brackets. The fact that the 308 periodicity is an alias of the P2 period is also particularly apparent in the fact that a peak at this period appears in Figures 2(b) and 3(b) also in the lower curves. These are the PSa of our synthetic models for the Yz and the Ya LCs, in which the 308 period has not been planted and therefore the signal at this period must be an artifact.

The period 339 d is also suspected to be a mere alias due to the yearly cycle of the observations. The peak corresponding to this period in the upper curve in Figures 2(b), 3(b) and 4(b) is marked with the number in brackets. As seen in the lower curves in these figures, there is a signal around this periodicity also in our corresponding synthetic LCs, in which the 339 d period has not been included.   It is therefore very likely that the 339 d feature in the PS of the real data is also due to the yearly cycle in the observations.

The 339 periodicity in the optical LC may well be a beat of the period of 365 days with  the P1/4=4750 d periodicity, which is rather pronounced in the LC of the star: 
$ 1/(1/365+4/19000) = 338.95 $

\subsubsection{The 161 d periodicity}

Frackowiak et al. (2003) identified  a 161.3 d periodicity in the NIR LC of the system that they have analyzed. They have attributes this period to pulsations of the giant. Their LC extends over the period 1990-1999, about 40 percent of the duration of our Yb LC. We do not find a significant signal at the frequency corresponding to 161 d period. A very week feature at this frequency can be seen in the PS of a partial section of our Yb function covering the 1990-1999 time interval. One has to bear in mind, though, that Frackowiak et al. (2003) found the signal in the IR band while we are analyzing the visible light of the star.

\section{Discussion}

\subsection{The pacemakers} 

Cyclic variations in the brightness of MWC 560 are known already for quite some time as major components of the LC of this variable star (Doroshenko et al. (1993), Gromadzki et al. (2007)). In this work we show that the major variations in the LC of the star during the last 87 years may well be interpreted as manifestations of 3 basic periodic oscillations, together with a few  harmonics of one of them and some beats between pairs among them. 

The question what is the nature of the clocks responsible for these 3 periods is now naturally suggesting itself.

It has been suggested that the MWC 560 binary system is seen at very low inclination, possibly close to a pole on situation. This could be the reason why so far there are no spectroscopic observations that enable determination with certainty what is the binary orbital period of the system. It has been suggested by a few investigators, that P2=1943 is the one (Doroshenko et al. (1993), Gromadzki et al. (2007)), but so far it seems that there is no compelling proof for such an identification.

The ignorance, or at least the uncertainty, in the knowledge of such a fundamental parameter as the binary period of the system, makes it even harder to determine the nature of the 3 distinguishable periods that are identified in the data. 

At this level of our understanding of the MWC 560 stellar system we can discuss the findings of this paper only in qualitative terms, by taking into account the order of magnitude of the 3 periodicities and by possible analogies with other multi-periodic symbiotic systems for which the nature of the major cycles operating in the system may be slightly better understood.

\subsection{The 19000 d periodicity}

On account of its length alone we suggest that the P1 periodicity is that of a Solar-like magnetic dynamo operating in the outer layers of the cool component of the system.

The length of the P1 period is about twice the length of the full 22 years cycle of the global magnetic field of the Sun. The time interval covered by all the observations in the star amounts to no more than  1.6 cycles of this periodicity therefore its persistence in the LC of the star must remain questionable. The reality of this period during the time interval of the observations is however suggested by the appearance in the LCs of the system of beats of this periodicity with the P2 or the P4 periods, as discussed in section 4.4. Also the observed LC includes 3.3 cycles of the P1/2=9500 d periodicity, the effect of which on the LC is undeniable as demonstrated in Figure 2(c).

The  P1/2= $\sim$26 years period is an analogue of the 11 years magnetic cycle of the Sun. The brightness of the star is correlated with this period since as in the Sun, the mass loss rate and hence the accretion rate onto the hot component environment is modulated by the surface magnetic variability (see, for example Schwenn (2006) for a review of the complex interrelation between solar cycle and solar wind).  A magnetic cycle of this length is not uncommon in stars. For example, Olah et al. (2009) found in the star V833 Tau, a magnetic cycle of  27-30 years. This K5V star with a rotation period of 1.788 d is admittedly  rather different from our M5III giant rotating with a period of hundreds days but it demonstrates that magnetic cycles of a time scale of tens of years are not uncommon in stellar physics.

The peak in the PS of the Yz LC representing the full P1 period, is due to the fact that the two successive maxima of the P1/2 cycle within one cycle of the P1 period are different in their amplitude 

Recently, Hric et al. (2014) have raised some doubts regarding our suggestion of the operation of a solar type magnetic cycle in the outer layers of the giant star in the AG Dra symbiotic system, in an analogy to our suggestion here. We do share their doubts mainly for the reason given by these authors, namely, the present lack of a direct evidence for a magnetic field or of magnetic activity in the atmosphere of the cool components of symbiotic stars. However, the lack of evidence is certainly not an evidence for the lacking. Direct detection of the operation of magnetic cycle in stars other than the Sun is a difficult undertaking. For one thing, it requires a very long time base for any type of observations. A major tool for detecting an operation of a magnetic cycle is an intensive monitoring of the emission components at the center of the CaII H \& K lines in the spectrum of the star (Wilson 1978). In the complex absorption and emission spectrum of symbiotics, these lines can hardly be observed, if at all, much less their intensity can be measured. 

The case of the Sun shows, as described in details in the Schwenn (2007) review, that the magnetic cycle may modify extensively the characteristics of the solar wind through its dramatic effect on local fields, without affecting drastically the intensity of the global magnetic field of the star. In symbiotic systems such as AG Dra and MWC 560, variations in the rate and the dynamics of the giant stellar wind, if indeed driven by solar type magnetic dynamo, are translated into large variations in the optical brightness of the system due to the presence of the close WD neighbor and the accretion process onto its environment. Thus in fact, if our suggestion turns out to be a correct interpretation of the thousands day variability of symbiotics, these systems may serve as laboratories for studying the operation of magnetic cycles in the atmospheres of giant stars.

\subsection{The 1943 periodicity}

It is by now commonly assumed by most investigators of the star MWC 560 that the P2=1943 d period is that of the orbital revolution of this binary system. We tend to accept this assumption but not without some caution remembering that so far there is no observational compelling evidence that this is in fact the case. 

\subsection{The 722 d periodicity}

It should be noted that on the basis of the information at hand  the period P4 that we discover in the LC of MWC 560 may not be a fundamental periodicity of the system. A more fundamental one could be any one of the 6 beating periods of P4 with the P1 or P2 periods, as presented in Table 3 . In such a case the 1150 periodicity would be a beat period of that more fundamental one with the corresponding P1 or the P2 period. 

We suggest that this is indeed the case and the fundamental periodicity of the system is P3=722 d. The main reason for this identification lies in the fact that the signal of the P3 period is clearly present in all 3 observed LCs that we are analyzing. Furthermore, when we fit P3 by least squares to the Ya LC and independently also the Yb Lc, the two resulting waves have virtually the same phase. 

Gromadzki et al. (2007) have also found a period of 741 d in the PS that they analyzed, which is a partial section of our Yb function. They did however cautioned that being close to a two terrestrial years duration, this periodicity may be an artifact of the observational materials. 

In assessing this possibility we first note that along the 87 years of observations there is no obvious reason why a two years cycle, rather than 1 year, should leave a mark on the observed LC.

As a test for this possibility we fitted the two periods, 722 and 730 days to each of the 2 sections of the LC, independently of one another. Table 5 presents the amplitude of each wave in each of the two sections, as well as the difference in phase between the 2 waves. We see that P3 preserves its phase throughout the 87 years of the observations, while the 2 year cycle is incoherent. We consider this result as evidence for the reality of the 722 d cycle in the system.

\subsection{Qualitative Model}

As suggested above the brightness variations of the system with the P1 period and its harmonics is following the magnetic dynamo cycle in the outer layers of the giant. The magnetic activity modulates the mass loss rate from the giant and also its dynamics and hence of the accretion rate onto the hot component.

The modulation at the binary period P2 is primarily due to the eccentricity of the binary orbit. Zamanov et al. (2010) estimated it to be as high as 0.68. Close to periastron passages of the system, the L1 point of its equi-potential Roche lobes is getting closer to the giant center, exposing deeper layers of the giant atmosphere to the gravitational pool of the companion. At such phases of the orbital period, accretion onto the hot component assumes particular high rate, while at apastron, accretion occurs at lower rate.

We now suggest that the P3=722 d is the sidereal rotation period of the giant. Zamanov et al. (2010) estimated the rotation period of the giant as 144-360 d. While our suggestion is about twice their upper value it is qualitatively in agreement with their conclusion that the rotation of the giant is rather far from synchronization with the binary revolution. The weak variability of the system with the P3 period may accordingly be due to spots on the surface of the giant. For such an interpretation, the rotation axis of the giant must be inclined with respect to the line of sight, due to an inclination of the binary plane and/or an inclination of the giant rotation axis with respect to that plane. 
The second
The period P4=1150 d satisfies  P4=1/(1/P3-1/P2) (See Table 3), i.e. it is the synodic rotation period of the giant for a prograde rotating star. At a given location on the surface of the giant, for example, the environment of the North or the South pole of an oblique dipole magnetic field, or a big spot on the star surface, P4 is the time interval between two successive passages through that location of the tidal bulge in the atmosphere of the giant that is due to the gravitation pool of the companion. If a local magnetic field induces intense  mass ejection from the giant surface at its locality, as the case is for some components of the Solar wind (Schwenn 2006), its effect is mostly pronounced at the tidal bulge of the star where the stellar outer layers are gravitational less bound to the star. This could be the origin of a modulation of the mass accretion rate onto the hot component at the P4 periodicity, and hence to the P4 period in the LC of the star. 

It should be noted, however, that one cannot rule out the possibility that the 1150 d period is the fundamental one after all, being the sidereal rotation period of the giant. In that case, 722 d is the synodic rotation period of a retrograde rotating star.

\subsection{Coherence 2} 

In our analysis in Section 4.1 we consider two possibilities regarding the phases of the periodic variation before and after the large brightening event of spring 1989. We call case I the possibility that this event has not affected the periodic content of the LC. Case II denotes the possibility that the brightening event was accompanied by a change in the phase of one or more of the  periodic waves underlying the LC. We find, even under the analysis that allows the case II situation, that the 3 fundamental periods of the system, P1, P2 and P3 preserve the same phase before and after the 1989 event.

For the P4 periodicity we find a phase shift of 0.2 between Ya and Yb.  In Table 4 we see that for the high harmonics P1/5 and P1/6 there is also a non negligible phase difference between the 2 sections of the LC. We are unable to determine whether or not these phase differences are significant, namely, that they testify on some reset process of the pacemakers operating in the system that accompanied the 1989 event. Alternatively these phase differences may be due to contamination of a pure representation of these periodicities in the LC of the star by brightness fluctuations on smaller time scale and by noise in the observational data. The data at hand do not allow us to ascertain whether case I or case II prevail in the system.

\section{Summary}

We present in this paper a 87 years light curve of the symbiotic star MWC 560. We showed that most of its observed optical variability may be accounted for by the operation of 3 clocks in the system and some beats between pairs of their periods. The longest period P1=19000 d appears in the LC with its 6 lowest harmonics. Establishing that these 3 periods are the seeds of the historic oscillatory behavior of the star is the main result of this work. It should be noted, however, that no explanation is being suggested in this paper for the 1989 cataclysmic event, whereby the overall optical luminosity of the star suddenly tripled its value within a few dozen days.

In way of interpretation we suggest that P1 is the length of a solar-like magnetic dynamo cycle that operates in the outer layers of the giant member of the system. P2=1943 d is the orbital period of the binary system and P3=722 d is the sidereal rotation period of the giant. The optical broad band output of the system is correlated with the mass accretion rate from the giant onto the hot component environment, which in turn is modulated directly by the first two of these periods. The mass accretion rate is also modulated by the period  P4=1150 d which is the synodic rotation period of the giant, the period of rotation of the gravitational tidal bulge in the atmosphere of the giant, as measured by an observer at a fixed location on the surface of the rotating giant. As Table 3 shows, the varying accretion rate with the P4 periodicity is also modulated with the first and the second harmonics of the P1 cycle.

MWC 560 is joining a group of 6 other symbiotic stars with recorded LCs that cover tens to over hundred years, the data of which we have assembled from the literature and from the invaluable treasures of AAVSO. We found in the LCs of all 7 stars traces of  similar cycles that affect in similar ways their optical luminosity. Table 6 presents the value of some basic parameters that characterize the 7 members of this group of symbiotic stars. 

If in the next few years, no cataclysmic event, similar to the 1989  one, is taking place in the system,  by extrapolating into the future the synthetic LC  shown as solid curve in Figure 2(a), we can predict that the mean, over a few hundred days, of the visual brightness of the star will continue to rise, reaching a maximum value around the end of the year 2016.

\section{Acknowledgments}

We are indebted to the American Association of Variable Stars Observers and to its dedicated members for the use that we made in the treasures of their data archive, without which this work could not have been done.

\section*{References}

\def\ref{\par\noindent\hangindent 18pt}

\ref Dobrzycka, D. et al. 1996, AJ. 111,414
\ref Doroshenko, V.T.,  Goranskij, V.P., Efimov, Yu.S. 1993, IBVS, 3824
\ref Fekel, F.C., Joyce, R.R., Hinkle, K.H., et al. 2000a, AJ, 119, 1375
\ref Fekel, F.C., Hinkle, K.H., Joyce, R.R., et al. 2000b, AJ, 120, 325
\ref Frackowiak, S. M., et al. 2003, arXiv:astro-ph/0210447
\ref Gromadzki, M.,  et al. 2007, A\&A, 463, 703
\ref Formiggini, L. \& Leibowitz, E.M. 2006, MNRAS, 372, 1325
\ref Formiggini, L. \& Leibowitz, E.M. 2012, MNRAS, 422, 2648
\ref Hric, L., G\'{a}lis, R., Leedj\"{a}rv, L. , et al. 2014, MNRAS, 443, 1103 
\ref Iijima, T. 2002, A\&A, 391, 617 
\ref Leedj\"{a}rv, L. 2003 in Asp Conf. Ser. 303, Symbiotic Stars Probing Stellar Evolution, ed. R.L.M., Corradi, R., Mikolajewska \& T. J. Mahoney 
(San Francisco,CA:ASP), 433)
\ref Leibowitz, E.M., \& Formiggini, L. 2006, MNRAS, 366, 675
\ref Leibowitz, E.M., \& Formiggini, L. 2008, MNRAS, 385, 445
\ref Leibowitz, E.M., \& Formiggini, L.  2011, MNRAS, 414, 2406
\ref Leibowitz, E.M., \& Formiggini, L.  2013, AJ, 146, 117
\ref Luthardt,  R.  1991, IBVS 3563
\ref Merrill, P. W., \& Burwell, C.G. 1943, ApJ, 98, 153
\ref Michalitsianos, A.G., et al. 1991, ApJ, 371, 761
\ref Michalitsianos, A.G., et al. 1993, ApJ, 409, L53
\ref M\"{u}rset, U., \&  Schmid, H.M. 1999, A\&A Supp., 137, 473
\ref M\"{u}rset, U., Nussbaumer, H., Schmid, H.M. , et al. 1991, A\&A, 248, 458
\ref Olah, K., et al. 2009, A\&A 501,703
\ref Roberts, D.H., Lehar, J., Dreher, J.W. 1987, AJ, 93, 968
\ref Sanduleak, N., \& Stephenson, C. 1973, ApJ, 185, 899
\ref Scargle, J.D. 1982, ApJ, 263, 835
\ref Schmid, H. M., et al. 2001, A\&A, 377, 206
\ref Schwenn, R.  2006, Space Science Reviews,  124,  51
\ref Stoyanov, K., \& Zamanov, R. K. 2014, arXiv:1408.6050v1
\ref Tomov, T. 1990, IUA Circ. No. 4955
\ref Tomov, T., et al. 1996,  A\&A Supp., 116, 1
\ref Wilson, O. C. 1978, ApJ, 226, 379
\ref Zamanov, R. K., et al. 2010, AN., 331, 282
\ref Zamanov, R. K., et al. 2011, IBVS, 5995
\ref Zhu, Z.X., Friedjung, M., Zhao, G., et al. 1999, A\&A Supp, 140, 69

\clearpage

\begin{figure}
\epsscale{.50}
\includegraphics[angle=0,scale=.90]{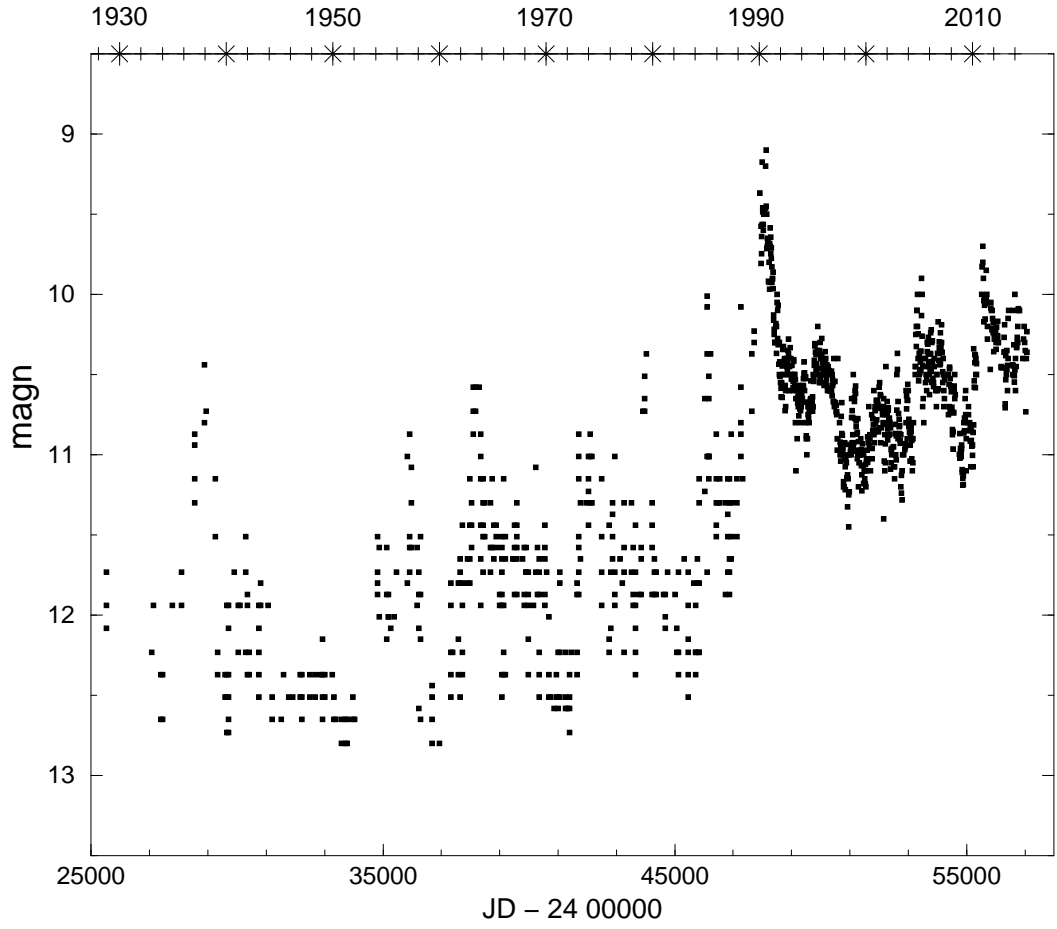}
\caption{The observed light curve (LC) of MWC 560 from September 1928 to February 2015.  \label{fig1}}.
\end{figure}

\clearpage

\begin{figure}
\epsscale{.50}
\includegraphics[angle=0,scale=.70]{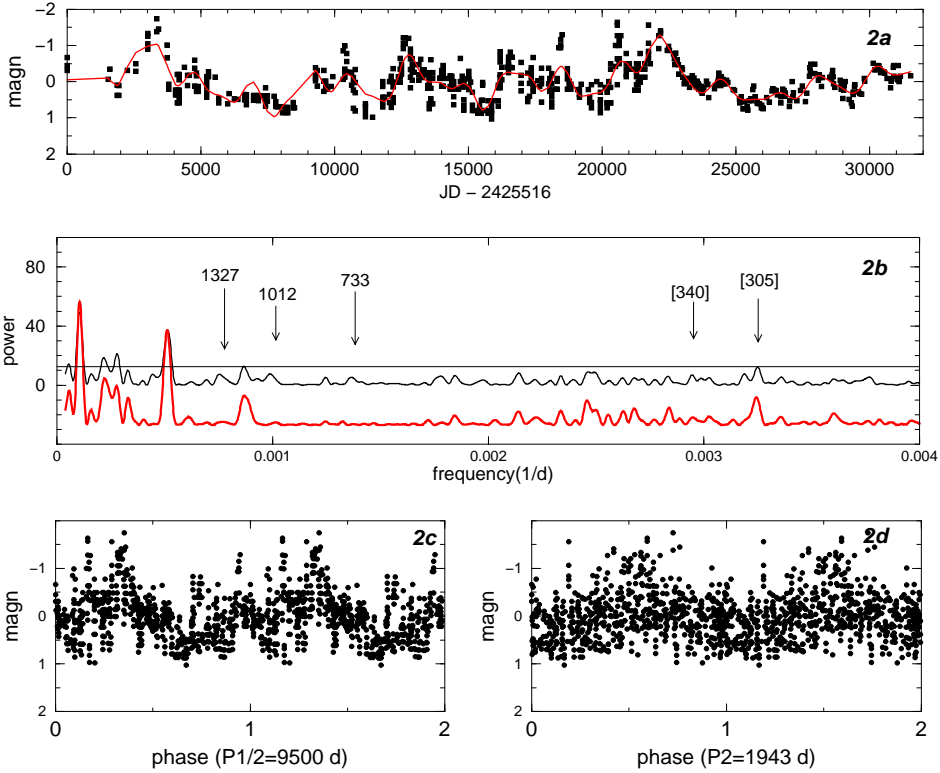}
\caption{(a)The detrended and diluted light curve (LC) of MWC 560 from September 1928 to February 2015. The line is the synthetic LC constructed with the periods P1, P2, P4 and six harmonics of P1 (see section 4.1). (b) Upper solid line is the power spectrum (PS) of the observed LC. Lower line is the PS of the synthetic LC. (c) The LC folded onto the period P1/2=9500 d. The cycle is exhibited twice. (d) The LC folded onto the period P2=1943 d.  \label{fig2}}.
\end{figure}

\clearpage
\begin{figure}
\includegraphics[angle=0,scale=.80]{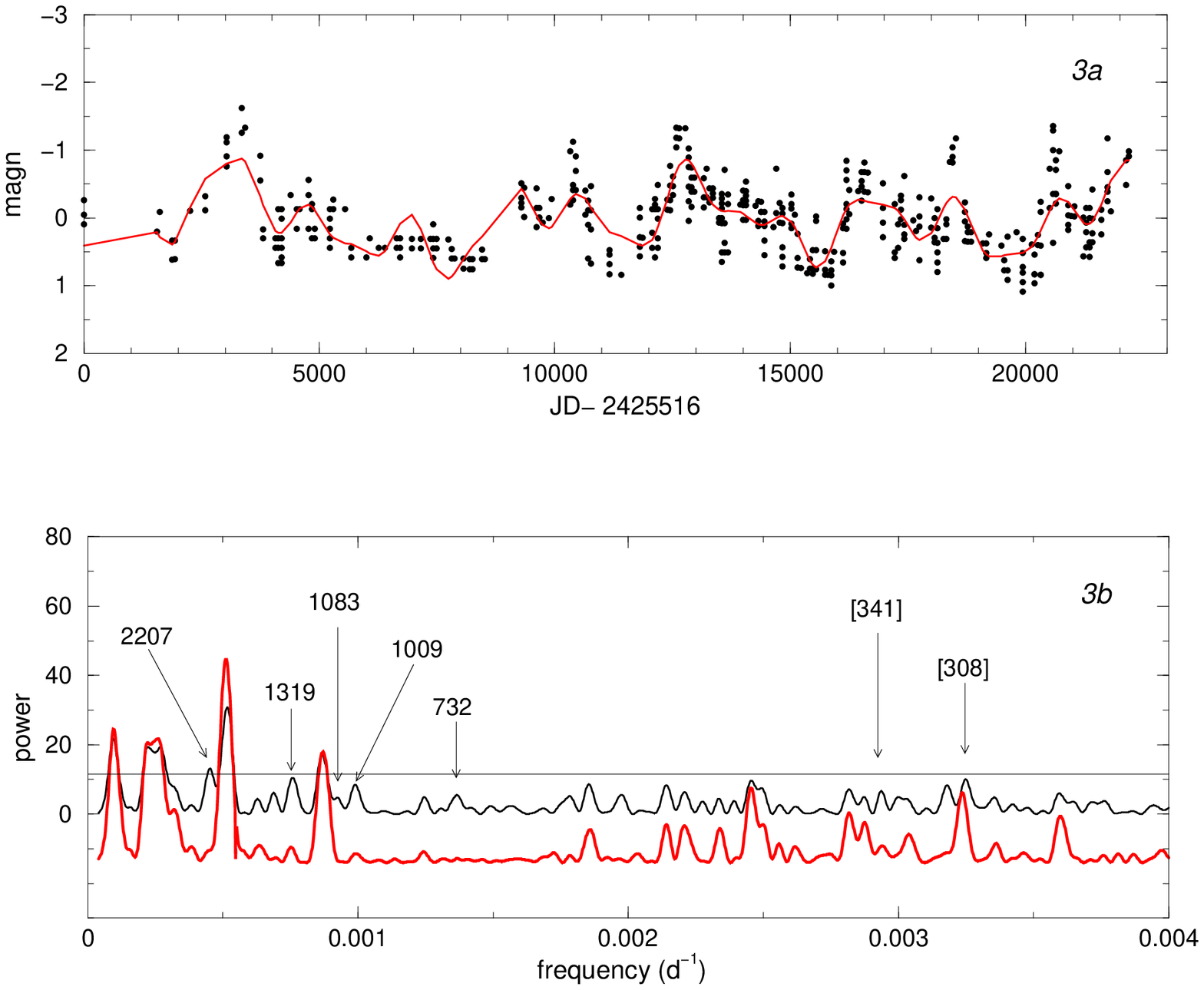}

\caption{(a) Points are the observed LC of section A. The line is the synthetic LC. (b) Upper solid line is the power spectrum (PS) of the observed LC of section A presented in frame a. Lower line is the PS of the synthetic LC.  \label{fig3}}. 

\end{figure}

\clearpage
\begin{figure}
\includegraphics[angle=0,scale=.80]{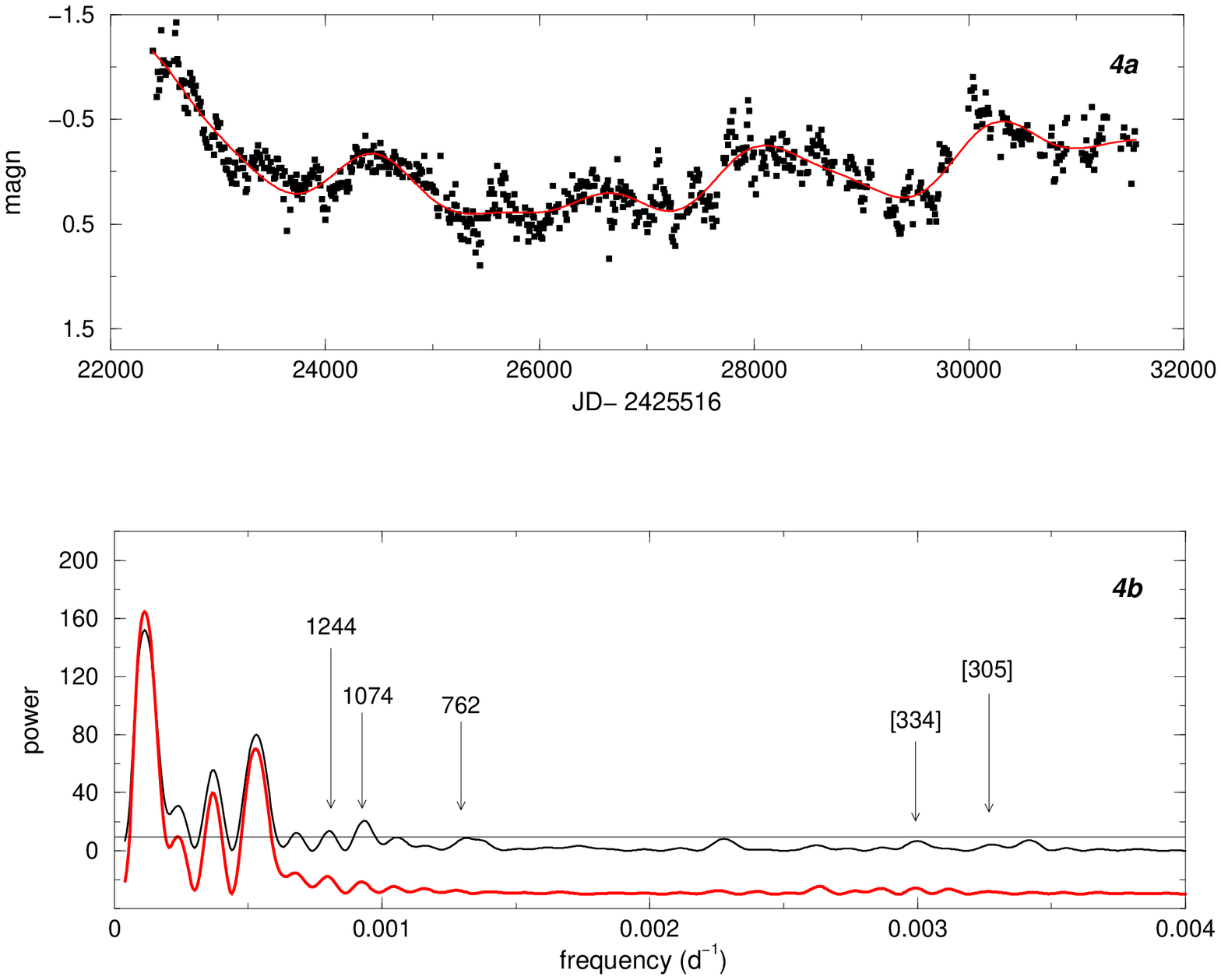}
\caption{(a) Points are the observed LC of section B. The line is the synthetic LC. (b) Upper solid line is the power spectrum (PS) of the observed LC of section B presented in frame a. Lower line is the PS of synthetic LC. \label{fig4}}. 

\end{figure}

\clearpage

\begin{deluxetable}{ccccccc}
\tablecolumns{7}
\tablecaption{The standard deviation of the difference between theoretical and observed LCs and PSa}
\tablewidth{0pt}
\tablehead{
\colhead{St. Dev.} & \colhead{Yta} & \colhead{Ysa} & \colhead{Ytb} & \colhead{Ysb} &
\colhead{YtzI} & \colhead{YszI}
}

\startdata
S$_{lc}$  &  0.348 & 0.353 & 0.161 & 0.172 & 0.339 & 0.332 \\
S$_{ps}$  & 3.343   & 3.276  &  4.179  & 4.705  & 3.893 & 3.332\\

\enddata
\end{deluxetable}

\begin{deluxetable}{cccc}
\tablecolumns{4}
\tablecaption{The standard deviation of the difference between theoretical and observed LCs and PSa}
\tablewidth{0pt}
\tablehead{
\colhead{St. Dev.} & \colhead{ YtzII} & \colhead{ YfzII}  & \colhead{YszII} 
}
\startdata
 S$_{lc}$ & 0.305  &  0.320 &  0.311 \\
 S$_{PS}$ &  3.138 &   3.446  &   2.923 \\
\enddata
\end{deluxetable}

\clearpage

\begin{deluxetable}{cccccc}
\tablecolumns{5}
\tablecaption{All beats between pairs of periods among P1, P2 and P4 periodicities. In the last three columns are periods of distinguishable peaks in the PS of the LC of the respective columns}
\tablewidth{0pt}
\tablehead{
\colhead{Beats period }& \colhead{period } & \colhead{ Ya  peaks} & \colhead{ Yb peaks} & \colhead{Yz peaks} 
}
\startdata
  1/(1/P2-1/P1) & 2164 &   2207 &       &        \\
  1/(1/P2+1/P1) & 1763 &        &       &        \\
  1/(1/P4-1/P1) & 1224 &        & 1244  &         \\
  1/(1/P4+1/P1) & 1084 &   1083 &  1074 &         \\
  1/(1/P4+1/P2) &  722 &    732 &   762 &    733  \\
  1/(1/P4-1/P2) & 2818 &        &       &         \\
  1/(1/P4-2/P1) & 1308 &   1319 &       &   1327  \\
  1/(1/P4+2/P1) & 1026 &   1009 &       &   1012  \\
\enddata

\end{deluxetable}

\begin{deluxetable}{ccccc}
\tablecolumns{4}
\tablecaption{The differences in phase between waves fitted to Sec A and Sec B }
\tablewidth{0pt}
\tablehead{ 
\colhead{Model Period } & \colhead{ phase B - phase A} & \colhead{'clean' period} & \colhead{phase B - phase A} 
}
\startdata
           P1/2 = 9500  &    0.0080 &    9507 &   0.4785 \\
           P1/4 = 4750  &    0.0688  &   4664 &   0.2122 \\
           P1/5 = 3800  &    0.1927  &   3669 &  -0.4044 \\
           P1/6 = 3167  &   -0.2979  &  3107 &   0.0205 \\
           P2   =  1943 &    0.0019 &    1933 &    0.0634 \\
\enddata
\end{deluxetable}

\begin{deluxetable}{cccc}
\tablecolumns{4}
\tablecaption{The amplitudes of waves fitted to Sec A and Sec B and the phase difference between them
}
\tablewidth{0pt}
\tablehead{ 
\colhead{Period } & \colhead{ Amplitude A } & \colhead{ Amplitude B} & \colhead{phaseB-phaseA}
}
\startdata

  722 & 0.0800 & 0.0701 & 0.063 \\
  730 & 0.1110 & 0.0719 & 0.239\\ 

\enddata
\end{deluxetable}

\begin{deluxetable}{ccccccccccc}
\tabletypesize{\footnotesize}
\tablecolumns{11}
\tablecaption{Comparison of the properties of seven symbiotic systems.}
\tablewidth{0pt}
\tablehead{ 
\colhead{Name} &\colhead{ Sp.Ty}  & \colhead{  Ref.} & \colhead{Bin per}& \colhead{Ref.} & \colhead{Giant Spin per}& \colhead{Ref.}& \colhead{Tidal wave per}& \colhead{Ref.}& \colhead{Solar-type per}& \colhead{Ref.} 
} 
\startdata
Z And   & M4 III   & 1 & 759.0   & 5  & 482    & 8  &  1317  & 8 & 7550   & 8\\
BF Cyg  & M5 III   & 1 & 757.3   & 6  & 798.8  & 6  & 14580  & 6 & 5375   & 6 \\
YY Her  & M4 III   & 1 & 593.2   & 7  & 551.4  & 7  &  7825  & 7 & 4650   & 7 \\
BX Mon  & M5 III   & 1 & 1256    & 4  & 656    & 9  &  1373  & 9 & 7370   & 9 \\
AG Dra  & K2 II    & 3 &  548    & 5  & 1160   & 10 &  373.5 & 10 & 5300  & 10\\
AX Per  & M4.5 III & 1 & 681.48  & 11 &        &    &        &    &11586* & 11 \\
MWC 560 & M5.5 III & 1 & 1943    & 12 &  722   & 12 &  1150  & 12 & 9500  & 12 \\

\enddata

\tablenotetext{*}{This number is one half the period value presented in the
AX Per paper (11) where it was suggested that it is a pulsation period of
the star. However, as pointed out in that paper, it may also be the period
of a solar-like magnetic cycle.}
\tablerefs{ (1) Murset \& Schmid (1999), (2) Murset et al. (1991), 
(3) Zhu et al. (1999), (4) Fekel et al. (2000a), 
(5) Fekel et al. (2000b), (6) Leibowitz \& Formiggini (2006), 
(7) Formiggini \& Leibowitz (2006), 
(8) Leibowitz \& Formiggini (2008),(9) Leibowitz \& Formiggini (2011), 
(10) Formiggini \& Leibowitz (2012), 
(11) Leibowitz \& Formiggini (2013) and (12) this paper.}
\end{deluxetable}

\end{document}